\documentclass[10pt]{iopart}

\usepackage{graphicx}

\begin{document}

\title[Influence of defects on the CrAs, CrSe and CrSb alloys]{Influence of defects on the electronic and magnetic properties
of half-metallic CrAs, CrSe and CrSb zinc-blende compounds}

\author{I Galanakis and S G Pouliasis}

\address{Department of Materials Science, School of Natural
  Sciences, University of Patras, Patras 265 04, Greece}

 \ead{galanakis@upatras.gr}

\begin{abstract}
We present an extended study of single impurity atoms and atomic
swaps in half-metallic CrAs, CrSb and CrSe zinc-blende compounds.
Although the perfect alloys present a rather large gap in the
minority-spin band, all defects under study, with the exception of
void impurities at Cr and sp sites and Cr impurities at sp sites
(as long as no swap occurs), induce new states within the gap. The
Fermi level can be pinned within these new minority states
depending on the lattice constant used for the calculations and
the electronegativity of the sp atoms. Although these impurity
states are localized in space around the impurity atoms and very
fast we regain the bulk behavior, their interaction can lead to
wide bands within the gap and thus loss of the half-metallic
character.
\end{abstract}


\submitto{\JPD}

\maketitle

\twocolumn

\section{Introduction}\label{sec1}

The rapid emergence of the field of spintronics (also known as
magnetoelectronics \cite{Zutic}) brought to the center of
scientific research the so-called half-metallic ferromagnets (like
Heusler alloys \cite{deGroot,GalanakisHalf,GalanakisFull} or some
oxides \cite{Soulen}).  These compounds present metallic behavior
for one spin-band while they are semiconducting or insulators for
the other spin-band, resulting to perfect spin-polarization, at
least for the bulk, at the Fermi level. Except Heusler and oxides,
also transition-metal chalcogenides like CrAs or CrSb and
pnictides like CrSe are known to present half-metallic
ferromagnetism when they crystallize in the metastable zinc-blende
structure. The first experimental evidence was provided in the
case of CrAs thin-films by the group of Akinaga in 2000
\cite{Akinaga2000} and many more experiments have confirmed these
results \cite{experiments}. Experiments agree with prediction of
ab-initio calculations performed by several groups
\cite{MavropoulosZB,GalaZB,calculations,Shirai}. In the case of
the half-metallic ferromagnets like CrAs or CrSe, the gap in the
minority-spin band arises from the hybridization between the
p-states of the $sp$ atom and the triple-degenerated $t_{2g}$
states  of the transition-metal and as a result the total
spin-moment, $M_t$, follows the Slater-Pauling (SP) behavior being
equal in $\mu_B$ to $Z_t-8$ where $Z_t$ the total number of
valence electrons in the unit cell \cite{MavropoulosZB}. Recently
theoretical works have appeared attacking also some crucial
aspects of these alloys like  the exchange bias in
ferro-/antiferromagnetic interfaces \cite{Nakamura2006}, the
stability of the zinc-blende structure \cite{Xie2003}, the
dynamical correlations \cite{Chioncel2006}, the interfaces with
semiconductors \cite{Interfaces}, the exchange interaction
\cite{Sasioglu-Gala} and  the temperature effects
\cite{MavropoulosTemp}. An extended overview on the properties of
these alloys can be found in reference \cite{Review}.

In a recent publication \cite{Rapid}, it was shown using the
coherent potential approximation \cite{koepernik} that for several
compounds (like CrAs, CrSb and CrSe)  an excess of the transition
metal atoms leads to half-metallic ferrimagnetism \cite{Leuken};
Cr-impurities occupying sites occupied by the $sp$ atoms in the
perfect bulk compound couple antiferromagnetically to the existing
Cr atoms at the ideal sites and destroy ferromagnetism keeping the
half-metallic character of the parent compounds. Later in
reference \cite{JMMM} it was shown that this also true when
instead of Cr-impurities we introduce other transition metal atoms
like V and Mn.

\begin{figure}
\begin{center}
\includegraphics[scale=0.4]{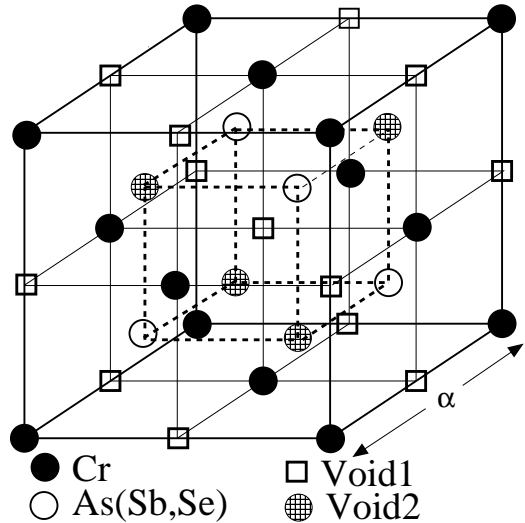}
\end{center} \caption{Schematic representation of the zinc-blende structure. To model
the system in our calculations we assume the existence of two
non-equivalent vacant sites. The lattice consists of 4 fcc
sublattices. The unit cell is that of an fcc lattice with four
atoms per unit cell: Cr at $(0\:0\:0)$, As(Sb or Se) at
$({1\over4}\:{1\over4}\:{1\over4})$ and the two vacant sites at
$({1\over2}\:{1\over2}\:{1\over2})$ (Void1) and
$({3\over4}\:{3\over4}\:{3\over4})$ (Void2).\label{figa}}
\end{figure}

\begin{table*}
\centering \caption{Total and atom-resolved spin magnetic moments
in $\mu_B$ for all four compounds under study. CrAs(GaAs)
corresponds to CrAs studied for the experimental lattice constant
of GaAs (0.565 nm) while (InAs) corresponds to the experimental
lattice constant of InAs of 0.606 nm.}
 \begin{tabular}{l|cccc} \hline \hline
& CrAs(GaAs) & CrAs(InAs)  &CrSb(InAs) & CrSe(InAs) \\ \hline
Cr & 3.017 & 3.267    &3.148 &   3.825 \\
 As(Sb,Se)& -0.198 & -0.382 & -0.249&   -0.103 \\
  Void1  & 0.005 & -0.029&
-0.036& 0.049 \\
Void2  & 0.122 &   0.080 & 0.084 & 0.162 \\
 TOTAL & 2.946 & 2.936 & 2.946 & 3.933  \\\hline \hline
\end{tabular}
\label{table1}
\end{table*}

CPA is an average method and does not provide details on the
nature of the single impurities. In this contribution we study in
detail the role of single impurities taking into account not only
the case of Cr antisites but all possible defects in half-metallic
zinc-blende compounds covering also the cases of atomic swaps
between  nearest neighbors. We have selected as test cases CrAs
which is the most widely studied alloy, CrSb which is isovalent to
CrAs and CrSe which has one electron more. We decided to use as
lattice constant the experimental one of InAs of 0.606nm. As it
was shown in reference \cite{MavropoulosZB} using the
Korringa-Kohn-Rostoker method (KKR) for this lattice constant all
three alloys are half-metals with the Fermi level exactly at the
middle of the gap and thus the effect of defect states is more
visible. The half-metallicity is reflected on the total
spin-moments which should be 3 $\mu_B$ for CrAs and CrSb and 4
$\mu_B$ for CrSe. As shown in table \ref{table1} the calculated
total spin moments are close to these values and small deviations
are due to the $\ell$-cutoff used in the calculations
\cite{GalanakisFull}. We have also decided to study a fourth case:
CrAs at the experimental lattice constant of GaAs (0.565 nm) which
is a usual substrate for thin films of this alloy. The latter one
is also half-metallic with the Fermi level near the right edge of
the gap.

\begin{table*}
\centering \caption{Structure of the impurity cluster which in
total contains 65 atoms. In parenthesis the distance between the
ith neighbors and the center atom in units of the lattice
constant.}
 \begin{tabular}{l|c|c|c|c} \hline \hline
Center  &  Cr  & As &  Void1  & Void2 \\

1st  (0.433013) &  4 As, 4 Void2 & 4 Cr, 4 Void1& 4 As,
4 Void2& 4 Cr, 4 Void1 \\

2nd  (0.500000)  & 6 Void1& 6 Void2& 6 Cr & 6 As \\
3rd  (0.707107) & 12 Cr & 12 As& 12 Void1 & 12 Void2\\
4th  (0.829156) & 12 As, 12 Void2 & 12 Cr, 12 Void1 & 12
As, 12 Void2 & 12 Cr, 12 Void1 \\
5th  (0.866025) & 8 Void1& 8 Void2& 8 Cr& 8 As \\
6th  (1.000000) & 6 Cr & 6 As& 6 Void1& 6 Void2 \\ \hline \hline
\end{tabular}
\label{table2}
\end{table*}

To perform our impurity calculations we have used a special
implementation of the KKR method  \cite{imp1,imp2}. First, a
self-consistent calculation is performed for the perfect compound
and the Green function is calculated and stored. Then we consider
an impurity cluster embedded in the perfect crystal. We use the
calculated Green function for the infinite host crystal and we
calculate self-consistently the electronic structure of the
impurity cluster in the real space using the Dyson equation. In
our case we found that an impurity cluster of 65 atoms is enough
to converge the spin magnetic moment of single impurity atoms at
the center of the cluster and atomic swaps between nearest
neighboring atoms, but not to calculate also the energetics of the
defects.  Our available computer sources did not allow us to
converge also the energetics with the size of the cluster. Within
the impurity cluster the impurity states survived up to the second
neighbors and  in all cases we regained the perfect crystal
behavior after the third beighbors.

The structure of the perfect crystal is shown in figure
\ref{figa}. The lattice is that of a fcc with four sites per unit
cell. We take into account two vacant sites to represent
adequately the zinc-blende structure. In other intermetallic
compounds like the full-Heusler alloys these two sites are
occupied \cite{GalanakisFull}. We assign the names Void1 and Void2
to the two vacant sites in order to study the case of Cr and sp
impurities at these sites. If we ignore the different chemical
elements the lattice is in reality a bcc one. Each Cr atom is at
the center of  a cube with four sp atoms and four Void2 sites as
first neighbors. The Void1 sites have the same local environment
as the Cr sites rotated by 90$^o$ degrees. Similarly each sp atom
and Void2 site are at a center of a cube with four Cr atoms and
four Void1 sites ar nearest neighbors. When we create the impurity
cluster we choose one of the four sites -Cr, sp atom, Void1 and
Void2- at the center of the cluster and in table \ref{table2} we
present the structure of the impurity cluster for all four cases.
We can seen that the clusters with Cr and Void1 sites at their
center are not identical since they have the same 1st and 4th
neighbor only which involve the sp and Void2 sites. The same
occurs for the clusters with a sp atom or a Void2 site at their
center. When we create an impurity we substitute the atom at the
center of the cluster and recalculate the electronic and magnetic
properties of the cluster. Similarly when we create an atomic swap
we exchange the atom at the center of the cluster with one of each
nearest neighbors.

In section \ref{sec2} we present the case of Cr impurities at
nearest neighboring sp or Void2 sites and in section \ref{sec3}
the case of sp  impurities at nearest neighboring Cr or Void1
sites. In section \ref{sec4} we discuss the other possible single
impuritites and in section \ref{sec5} we discuss the case of
atomic swaps between nearest neighboring atoms. Finally in section
\ref{sec6} we summarize and conclude. We should also mention here
that in table \ref{table1} we present the atom-resolved spin
moments for all four perfect bulk compounds under study to use
them as a reference when discussing the spin magnetic moment of
the impurities atoms (see reference \cite{MavropoulosZB} for an
extended discussion of their behavior). Moreover to distinguish
the two cases of CrAs we use the notation "(GaAs)" for the case of
the GaAs experimental lattice constant and "(InAs)" for the InAs
experimental lattice constant.

\begin{figure*}
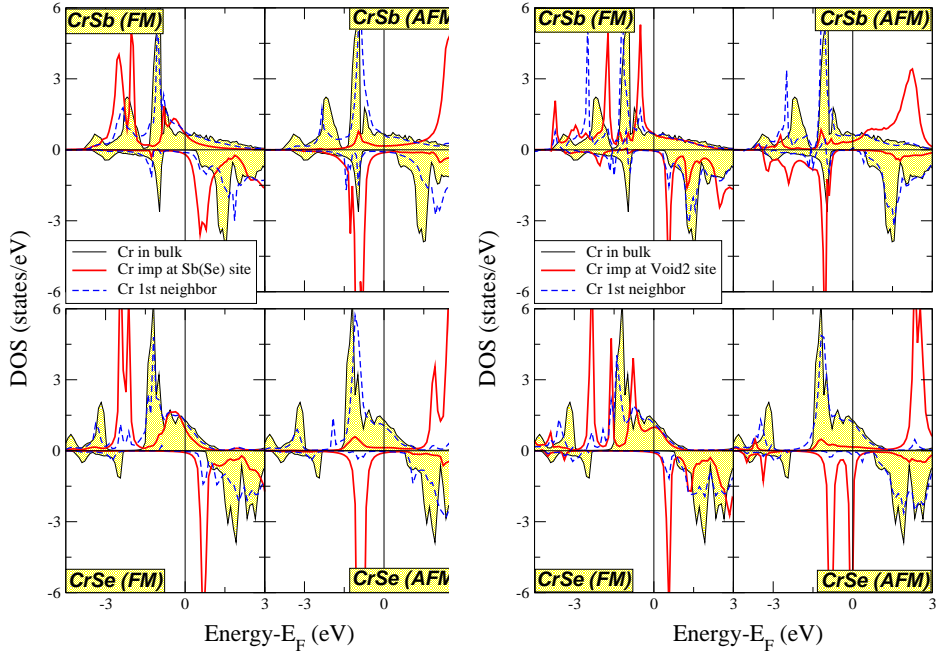

\includegraphics[scale=0.35]{ios1a.eps}
\includegraphics[scale=0.35]{ios1b.eps}
\caption{(Color online) Cr-resolved DOS for the case of Cr
impurity atoms at Sb(Se) sites in the left panel and at Void2
sites in the right panel (solid red lines) and their Cr first
neighbors (dashed blue lines) with respect to the perfect bulk
case (solid black line with shaded region). We present both cases
of ferromagnetic (FM) and antiferromagnetic
 (AFM) coupling of the Cr impurity spin moment (see text for details). The Fermi energy
 has been set as the zero of the energy axis. Positive DOS values correspond to
 the spin-up (majority-spin) electrons and negative values to the spin-down (minority-spin) electrons.\label{fig1}}
\end{figure*}

\section{Cr impurities at nearest neighbors sites}\label{sec2}

We will start our discussion from the case of Cr single impurities
at nearest neighboring sites since in references \cite{Rapid} and
\cite{JMMM} it was shown that Cr impurities at sites occupied by
the sp atom in the perfect crystal lead to half-metallic
ferrimagnetism. The Cr impurity atom has four Cr atoms and four
Void1 sites as first neighbors. The distance is so short that the
antiferromagnetic (AFM) coupling is favored between the spin
magnetic moment of the impurity atom and its closest Cr neighbors.
This is not surprising since it is well-known that both Cr and Mn
atoms show AFM or ferromagnetic (FM) coupling of their spin
moments depending on the distance between the neighboring
transition metal atoms. In our study we have expanded it to cover
also the case when Cr impurities appear at Void2 sites where the
local environment is the same as for Cr impurities at sp atom
sites.

In both cases of single Cr impurities, we found that depending on
the starting potential we were able to converge to two different
states: an antiferromagnetic one as in CPA calculations and a FM
one. This can be seen in table \ref{table3} where we present the
spin magnetic moments of the Cr impurity atoms and their first
neighbors. A close look at the energies shows that the AFM state
is more favorable energetically with respect to the FM case (we
were not able to converge the energy difference with the size of
the cluster but the total energy difference was of the order of
some eV). We will discuss also the FM case since in real
situations this state can occur either due to the method used to
grow the samples or due to an external field. We should also note
here that although we have tried we were not able to get a FM
solution for the case of CrAs at the GaAs lattice constant. GaAs
lattice constant is considerably smaller with respect to the InAs
lattice constant and thus the distance between the Cr impurity
atom and its first neighbors become much smaller and the FM
solution becomes completely unstable. Thus the possibility of the
FM coupling to occur depends strongly on the lattice constant. For
the compounds under study their lattice constant is imposed by the
substrate used to grow the films and thus we can exclude the FM
coupling to appear by choosing a suitable substrate.

\begin{table*}
\centering \caption{Atom-resolved spin magnetic moments for the
case of Cr impurity atoms at As(Sb or Se) and Void2 sites, and the
case of Cr-As(Sb or Se) and Cr-Void2 atomic swaps. "imp" stands
for impurity and "nn" stands for nearest neighbor atoms (1st
neighbors). We present results for both cases of coupling of the
Cr impurity spin moment with respect to the spin moments of the
other Cr atoms (ferromagnetic-FM and antiferromagnetic-AFM
cases).}
 \begin{tabular}{l|c|c|c|c|c|c|c} \hline \hline

& CrAs(GaAs) &   \multicolumn{2}{c|}{CrAs(InAs)} &
\multicolumn{2}{c|}{CrSb(InAs)}& \multicolumn{2}{c}{CrSe(InAs)}
\\ &AFM & FM & AFM & FM &AFM & FM &AFM
\\\hline \multicolumn{8}{c}{Cr impurity at As(Sb or Se)
site} \\
Cr imp &  -2.896&  4.388&   -3.690&  4.504&  -3.718&  4.511&
-3.642\\
 Cr nn  &  2.985&  3.602&  3.364&  3.537 & 3.301&
3.994&  3.715\\
\multicolumn{8}{c}{Cr impurity at Void2 site}
\\
Cr imp &  -1.129&    3.901&     -2.908&    3.854&  -2.689& 4.354&
-3.350 \\
 Cr nn  &  2.779&  3.419&  3.267 & 3.316&
3.175&  3.941&     3.703 \\
 \hline
\multicolumn{8}{c}{Cr-As(Sb or Se) atomic swaps} \\

Cr imp &  -2.840&    4.098&     -3.496&    4.204&  -3.558& 4.037&
-3.405 \\ sp imp  & 0.118&     0.075 & 0.162&  -0.326&
-0.325&    -0.055&    0.283\\

\multicolumn{8}{c}{Cr-Void2 atomic swaps} \\

Cr imp &  -2.224&    3.851&     -3.240&    3.787&  -3.157& 4.241&
-3.549\\ Void2 imp  &  -0.045 & -0.060&  -0.098&  -0.057& -0.077&
0.024&  -0.027\\

\hline \hline
\end{tabular}
\label{table3}
\end{table*}

\begin{figure}
\includegraphics[scale=0.35]{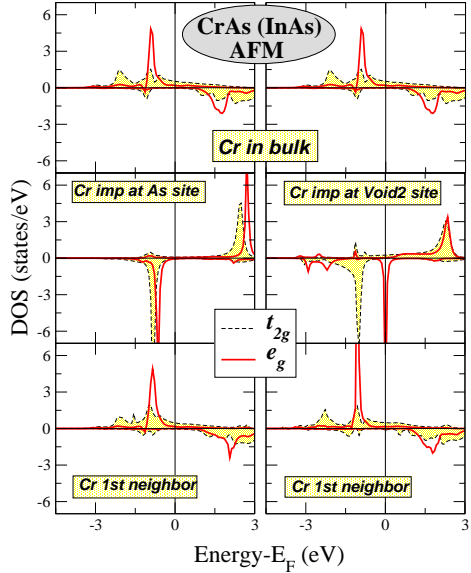}
\caption{(Color online) Upper panel: Cr $d$-DOS decomposed on the
$e_g$ and $t_{2g}$ orbitals for the bulk CrAs alloy at the InAs
lattice constant; middle and lower panel: same for the case of AFM
coupling of Cr impurity atoms at As (left panel) and Void2 (right
panel) sites and their first Cr neighbors.  \label{fig2}}
\end{figure}

First, we will shortly discuss the FM case. In figure \ref{fig2}
we have plotted the DOS of the Cr impurity atoms at Sb(Se) and
Void2 sites and their first Cr neighbors with respect to  the Cr
DOS in the perfect bulk CrSb and CrSe alloys (CrAs shows similar
behavior). The Cr impurity atoms have now four Cr atoms as first
neighbors instead of four Sb or Se atoms. The d electrons of the
Cr impurity atoms now can hybridize with the d electrons of the
neighboring atoms instead of the p electrons of the sp atom (only
the triple degenerated $t_{2g}$ electrons of Cr can couple to the
p orbital of the sp atom). The majority-spin electrons of the Cr
impurity are fixed in energy range since they form a common band
with the other Cr atoms but the center of the majority band is
shifted lower in energy. This leads to a shift of the unoccupied
minority-spin states lower in energy, since the exchange splitting
is comparable for all Cr atoms, and they become more localized in
energy. The small weight of the occupied minority electrons
vanishes since it was due to the hybridization with the p states
of the sp atom. The shift the unoccupied minority band is such
that the gap is scarcely affected and the half-metallicity is not
destroyed. The Cr first neighbors show a DOS very similar to the
bulk case since they have one Cr and three sp atoms as first
neighbors and the effect of the Cr impurity atom on their
properties is weak. There is no significant difference whether the
Cr impurity atom is located at the sp or Void2 site.  Majority
states are identical. In the minority-spin band the $e_g$ states
are located bear the Fermi level (the large pick just above the
Fermi level). In the case of Cr impurity atoms at the sp sites the
minority $t_{2g}$ states overlap with the $e_g$ states while in
the case of Cr impurity atoms at the Void2 sites the unoccupied
minority $t_{2g}$ states are slightly higher in energy. The spin
moments presented in table \ref{table1} do not show any unexpected
behavior. The Cr impurity atoms have a larger spin moment with
respect to the Cr atoms in the bulk case since the weight of the
minority occupied states has vanished as a result of the
hybridization with its first neighbors. The Cr atoms which are
first neighbors of the impurity atoms present spin moments almost
identical to the perfect compounds in table \ref{table1}.

The second and most interesting case is the AFM coupling between
the Cr impurity atom and the spin magnetic moments of  its nearest
Cr neighbors since this is susceptible of leading to half metallic
ferrimagnetism. This is crucial for applications since
ferrimagnets create smaller external fields and thus exhibit
smaller energy losses with respect to ferromagnets. In figure
\ref{fig1} we present also the cases of AFM coupling for Cr
impurity at Sb(Se) and Void2 sites in CrSb and CrSe alloys. In the
case when the Cr impurity atom is located at the sp atom site
there is a very intense pick in the minority spin band which is
occupied. This pick contains the 5 occupied bonding $d$ states of
the Cr impurity atom. But the spin magnetic moment of the impurity
atom is not -5 $\mu_B$ but around -3 to -3.7 $\mu_B$. This is due
to the majority $t_{2g}$ bands which are very extended in energy
and which are partially occupied due to their hybridization with
the $t_{2g}$ bands of the nearest neighboring $d$ bands.

When the Cr impurity atom is located at the Void2 site the
situation is more complex. While for CrSb the situation is similar
to the case of the Cr impurity at the Sb atom, for the other three
compounds there is now a double pick structure in the spin-down
band and the Fermi level falls within the second pick. This is
also reflected on the spin moments of the Cr impurity atom, the
absolute values of which are decreased with respect to the case of
the Cr-impurity at the sp atom site (in the case of CrAs in the
GaAs lattice constant the decrease is larger since also the weight
of the spin-up states increases slightly). The magnitude of the
decrease depends on the exact position of the Fermi level within
the pick. To elucidate this behavior we have plotted in figure
\ref{fig2} the projection of the d states on the
triple-degenerated $t_{2g}$ electrons and on the
double-degenerated $e_g$ states for the case of CrAs in the InAs
lattice constant. The $t_{2g}$ states are more delocalized and
couple to the p states of the sp atoms and thus are more extended
in energy. On the other hand the $e_g$ states can not couple to
the p states of the sp atoms and are more localized in energy.
This is reflected on the DOS of the Cr atom in the perfect crystal
presented in the upper panel of figure \ref{fig2} where the
$t_{2g}$ are more extended in energy. When we create the impurity
at the As site we see than in the minority spin band the $t_{2g}$
and $e_g$ states almost overlap, while in the case of the Cr
impurity at the Void2 site the $e_g$ states move higher in energy
and they are now located at the Fermi level. There is no immediate
explanation for this behavior since it should be attributed to the
different structure of the impurity clusters and not to the
immediate environment of the impurity atoms. Cr impurities have in
both cases 4 Cr atoms and four Void sites as first neighbors (the
Cr neighboring atoms have the same DOS and spin moments for both
cases) but when the impurity is located at the As site, it has 12
As atoms as 3rd neighbors while when it is located at the Void2
site it has 6 As atoms as 2nd neighbors. This difference probably
drives the shift of the minority $e_g$ states. In the case of CrSb
the shift is smaller leading to a more narrow and more intense
pick in the minority spina band in the case of the Cr impurity at
the Void2 site with respect to the case of the Cr impurity at the
Sb atom.

\begin{figure}
\includegraphics[scale=0.35]{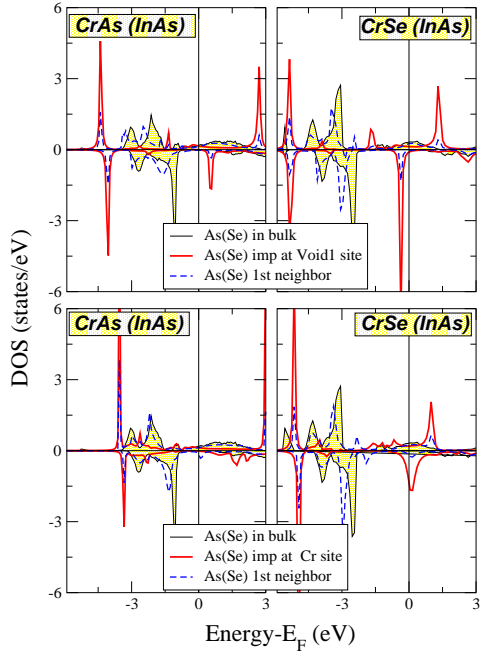}
\caption{(Color online) As(Se)-resolved DOS for the case of As(Se)
impurity atoms (solid red lines) and their 1st neighbors (dashed
blue lines) for the case of As(Se) impurities at Void1 sites in
the upper panel and Cr sites in the lower panel with respect to
the perfect bulk case (solid black line with shaded region).
\label{fig3}}
\end{figure}

\section{sp impurities at nearest neighbors sites}\label{sec3}

We will continue our study with the case of sp impurities at the
neighboring Cr and Void1 sites. The s electrons of the sp atoms
are located very low in energy, at around -10 eV, and play no role
to half-metallicity but this is not the case for the p electrons.
In figure \ref{fig3} we have plotted the DOS for both case of
As(Se) impurities at Void1 and Cr sites with respect to the bulk
CrAs and CrSe alloys. For both perfect bulk alloys the minority p
states of As(Se) are completely occupied leading to small negative
spin moments as can be seen in table \ref{table1}. When the As(Se)
atoms migrate to Cr or Void1 site they have four other As(Se)
atoms and four Void2 sites as nearest neighbors. Thus the p states
of the As(Se) impurity atoms have to hybridize with the p states
of the neighboring As(Se) atoms  which are almost completely
occupied instead of the Cr $t_2g$ states for which only the
majority states are occupied. This leads to an extended
reorganization of the p charge of the As(Se) impurity p states and
the charge is now mainly localized in intense narrow picks located
at around -3 eV and just below or above the gap. In the case of As
impurities at Cr sites the unoccupied spin-down states are well
above the Fermi level but the neighboring As sites show a small
spin-down pick pinned exactly at the Fermi level. When the As
impurity atom occurs at a Void1 site the difference in the
impurity cluster leads to a narrow spin down pick just above the
Fermi level. Thus the half-metallicity is preserved although the
gap shrinks. In the case of CrSe, Se atoms have one electron more
than As. When we create the impurity atom, its charge is no more
accommodated in transition-metal bands \cite{MavropoulosZB}
leading to an extra minority pick. When the Se impurity is located
at the Void1 site the Fermi level is just above the gap while when
the Se impurity is located at the Cr site the Fermi level falls
within this pick completely destroying half-metallicity. But our
results are for the InAs lattice constant. If we contract the
lattice constant we push the p states higher in energy and when we
expand it we push them lower in energy \cite{MavropoulosZB} and
thus it is very probable that the Fermi level falls within these
picks completely destroying the half-metallic character of the
parent compounds. The As(Se) nearest neighbors are affected by the
As(Se) impurity since now they have three Cr atoms and one sp atom
as nearest neighbors and  a small faction of their p charge moves
to the same energy region with the  picks of the impurity atom.

\begin{table*}
\centering \caption{Atom-resolved spin magnetic moments for
several cases of impurities. "imp" stands for impurity, "nn"
stands for nearest neighbor atoms (1st neighbors) and "nnn" for
next-nearest neighbors (2nd neighbors). In the last part on the
As(Sb,Se)-Void1 atomic swaps we present the minimum and maximum
values of Cr nn spin moments (see text for explanation).}
 \begin{tabular}{l|c|c|c|c} \hline \hline
& CrAs(GaAs) & CrAs(InAs) & CrSb(InAs) & CrSe(InAs) \\
\hline
\multicolumn{5}{c}{As(Sb,Se) impurity at Cr site} \\
sp imp  & 0.0627&     0.133&     0.177&     0.226\\ As nn & -0.162
& -0.232&    -0.119&    -0.067 \\
\multicolumn{5}{c}{As(Sb,Se) impurity at Void1 site} \\
sp imp &   0.005&      0.146&      0.155&      -0.572 \\ As nn &
-0.140&     -0.233&     -0.119&     -0.210\\ Cr nnn &  2.952&
3.439&   3.387&   3.717 \\
\multicolumn{5}{c}{Cr impurity at Void1 site} \\
Cr imp &  3.038&     3.505&     3.441&     3.993 \\
 As nn&-0.206&  -0.295&    -0.205&    -0.187\\
  Cr nnn  &  2.968 &    3.433&
3.338&  3.910 \\
\multicolumn{5}{c}{As(Sb,Se) impurity at Void2 site} \\
sp imp  &  -0.169&     0.455&      0.374&      -0.430\\ Cr nn&
2.583&   3.408&      3.322&      3.595 \\ As nnn &  -0.200&
-0.247& -0.166&   -0.161 \\
\multicolumn{5}{c}{Void impurity at Cr site} \\
Void imp &    -0.065&    -0.110 &   -0.102&   -0.019\\ As nn&
-0.314&    -0.569&    -0.353&    -0.233 \\
\multicolumn{5}{c}{Void impurity at As(Sb or Se) site} \\
Void imp &     0.367&     0.296&     0.285&     0.355\\ Cr nn &
3.502&     3.731&     3. 576&    4.063 \\
\multicolumn{5}{c}{As(Sb or Se)-Void1 atomic swaps} \\
sp imp&   -0.027&    0.156&     -0.232&    -0.469 \\ Void1 imp&
0.264&     0.262&     0.178&     0.232\\ Cr nn &  3.195 &
 3.737 &     3.579&      3.854
 \\& 3.532 & 3.777&  3.603&  4.059 \\
\hline \hline
\end{tabular}
\label{table4}
\end{table*}

We should also discuss briefly the spin moments presented in table
\ref{table4}. The impurity atoms have positive spin moments
contrary to the sp atoms at the perfect bulk compounds where they
have negative spin moments. This is due to the difference in the
hybridization, as we referred to above, which leads to occupancy
of antibonding p spin-up states. The only exception is the Se
impurity atom at the Vois1 site which has a negative spin moment.
The sp atoms which are the nearest neighbors of the impurity have
spin moments very close to the bulk case since the DOS are also
similar. Thus no general rule or conclusion can be drawn for the
case of sp atom impurities at Void1 and Cr sites since depending
on the chemical type of the sp atom and the lattice constant the
Fermi level can fall within a minority spn pick destroying the
half-metallicity.

\section{Other  impurities}\label{sec4}

\begin{figure}
\includegraphics[scale=0.35]{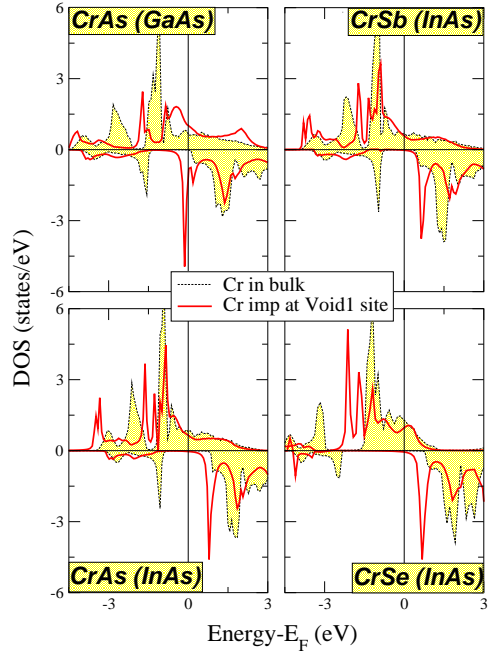}
\caption{(Color online) Cr-resolved DOS for the case of Cr
impurity atoms at Void1 sites (solid red lines) for all four
compounds under study with respect to the perfect bulk case
(dashed black line with shaded region). \label{fig4}}
\end{figure}

To conclude our study on single impurities we will also present
our results when the Cr or sp impurities appear at next-nearest
neighbors sites. When these impurities are created the impurity
atoms keep the same local environment as in the bulk compound
(same nearest neighbors) since as we have mentioned in section
\ref{sec1} Cr and Voi1 sites have 4 sp and 4 Void2 sites as
nearest neighbors (sp and Void2 sites have 4 Cr and 4 Void1 site
as nearest neighbors). In figure \ref{fig4} we present the DOS for
the case of  Cr impurity atoms at Void1 sites for all four
compounds with respect to the Cr atoms in the perfect compounds.
Although the local environment remains unchanged, the impurity
cluster as can be seen in table \ref{table2} is different. This
leads again to a shift of the $e_g$ minority spin states towards
lower energies. While in the perfect compounds the $e_g$ and
$t_{2g}$ minority states overlap, in the case of impurities the
$e_g$ states create a new pick just below the $t_{2g}$ states. For
the compounds at the InAs lattice constant the presence of the
pick leads to a shrinking of the gap but the half-metallicity is
preserved. But when we contract the lattice, as for CrAs at the
GaAs lattice constant, the Fermi level is located within the
minority $e_g$ pick since the Fermi level is already at the right
edge of the gap for the perfect bulk compound. The spin moments as
can be seen in table \ref{table4} are very close to the bulk
values and thus the no conclusion can be drawn from the spin
magnetic moments. Cr impurities at Void1 sites do not affect the
half metallicity only if the Fermi level is located at the
lower-energy edge or the middle of the gap.

\begin{figure}
\includegraphics[scale=0.35]{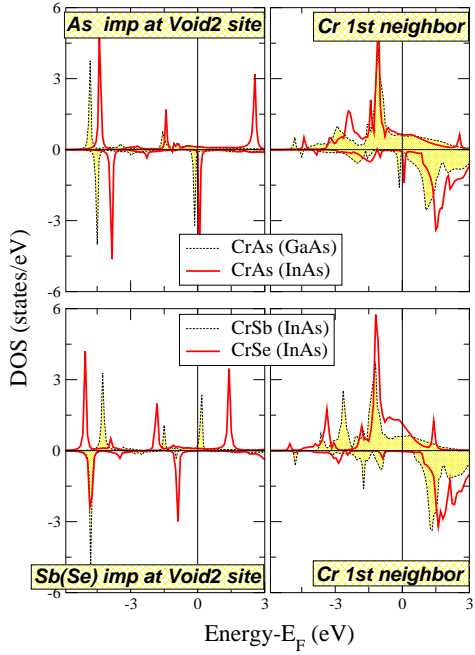}
\caption{(Color online) As(Sb or Se)-resolved DOS for the case of
As(Se) impurity atoms at Void2 sites and their 1st Cr neighbors.
\label{fig5}}
\end{figure}

We will now proceed with the case of sp impurities at Void2 sites.
We present for all four compounds the DOS of the sp impurity and
its closest Cr neighbor in figure \ref{fig5}. The situation is
also complex as in the case of sp impurities at Cr or Void1 sites.
The p states are again reorganized in narrow picks and the
half-metallic character depends on the exact position of these
picks with respect to the Fermi level. In the case of CrAs at GaAs
or InAs lattice constant, the Fermi level is located exactly just
below or just above such a minority spin pick. An image of the
states in this pick survives also for the Cr neighbors which now
have 5 instead of 4 sp atoms as nearest neighbors and the
hybridization effect for the $t_{2g}$ states is more intense
leading also to smaller Cr spin moments with respect to the bulk
case as can be seen in table \ref{table4}. In the case of CrSb and
CrSe these states are located lower in energy due to the larger
Coulomb repulsion between the p electrons (Sb is isovalent to As
but has also the 5d states occupied while Se has one electron more
than As) and the half-metallicity is preserved.

Finally we should also discuss the case of Void impurities at Cr
or sp sites. We do not present the DOS since it is almost
identical to the bulk cases for the neighboring atoms. This is
also reflected on the spin moments presented in table
\ref{table4}. When we create the Void impurity at the sp site, the
Cr nearest neighbors have now 3 instead of 4 neighboring sp atoms
and thus  the hybridization effect is smaller leading to slightly
larger Cr spin moments. The same occurs also when the Void
impurity is located at a Cr site and the absolute value of the
spin magnetic moment of the neighboring sp atoms increases
slightly. The width of the gap remains unchanged by the Void
impurities at Cr or sp sites and half-metallicity is not altered.

\begin{figure}
\includegraphics[scale=0.35]{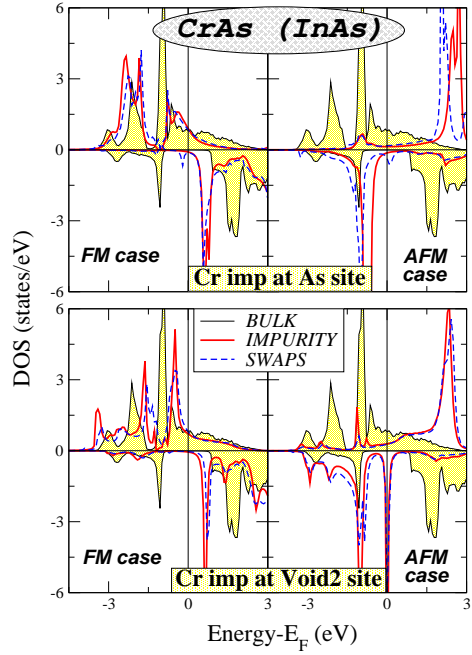}
\caption{(Color online) For the case of CrAs in the InAs lattice
constant, we present the Cr-DOS (for both AFM and FM cases) for
both Cr impurity atoms at As site (solid red line) and Cr-As
atomic swaps (dashed blue line) in the upper panel, and for both
Cr impurity atoms at Void2 sites (solid red line) and Cr-Void2
atomic swaps (dashed blue line) in the lower panel. With the solid
black line with the shaded region we present the perfect bulk
case. \label{fig6}}
\end{figure}

\section{Atomic swaps}\label{sec5}

In the last section we will present our results on the atomic
swaps. As we mentioned in section \ref{sec1} we performed
calculations for the case of atomic swaps between nearest
neighboring sites (i) Cr-sp atom, (ii) Cr-Void2, and (iii) sp
atom-Void1. Atomic swaps present properties very similar to the
properties of the single impurities. We will start our discussion
from the case of swaps involving Cr atoms. For both Cr-sp atom and
Cr-Void2 swaps we were able to converge two solutions for the InAs
lattice constant: a FM and an energetically-favorable AFM coupling
of the Cr impurity with respect to its neighboring atoms. For CrAs
at the GaAs lattice constant the distance between the Cr atoms is
very short to stabilize a FM solution and we were able to converge
only the AFM solution. In table \ref{table3} we present the spin
moments of the impurity atoms for all cases. In the case of swaps
each Cr impurity atom has 3 Cr atoms and one sp atom as nearest
neighbors while in the case of a Cr single impurity the
Cr-impurity atoms had 4 Cr atoms as first neighbors. Since the
$t_{2g}$ states of the Cr atoms hybridize strongly with the p
states of the sp atoms, this leads to an increased hybridization
in the case of the atomic swaps and thus to smaller absolute
values of the spin magnetic moments of the Cr atoms in both the FM
and AFM cases. In figure \ref{fig6} we present the Cr DOS for the
Cr impurity atoms at both As and Void2 sites for both FM and AFM
cases in the case of CrAs at the InAs lattice constant, and we
compare the perfect bulk case, the single impurity case discussed
in section \ref{sec2} and the swaps case. Overall the DOS of the
Cr impurity atom only scarcely changes between the single impurity
and swaps cases. In the case of the Cr impurity at the Void2 site
the $e_g$ states in the AFM solution are pinned at the Fermi level
destroying the half-metallicity as in the case of single impurity.
When the Cr impurity atom is located at the As site there is an
extra minority pick at the Fermi level in the case of atomic
swaps. To understand its origin we should keep in mind that also
an As atom moves at a Cr site in the case of swaps and as we can
see in the lower left panel of figure \ref{fig3} already in the
case of a single As impurity at a Cr site the neighboring atoms
present a small DOS at the Fermi level which destroyes the
half-metallicity. In the case of Cr-As swap, the As impurity atom
has one Cr atom and three As atoms as nearest neighbors and this
pick is present destroying the half-metallicity. Thus the
energetically favorable AFM case of Cr-As and Cr-Void2 swaps
induces states within the gap.

Finally we should also shortly discuss the third case of atomic
swaps: sp-Void1. In table \ref{table4} we present the spin
magnetic moments at the Void1 and sp impurity sites. The nearest
Cr atoms are no more equivalent and thus their total spin moment
varies between the values given in the table. We do not present
the DOS since results are similar to the case of single sp
impurities at Void1 sites presented in section \ref{sec3}. There
is a very narrow intense pick near the Fermi level as in the upper
panel of figure \ref{fig3} which can lead to the loss of
half-metallicity depending on its exact position with respect to
the Fermi level which is influenced from the electronegativity of
the sp atom and the lattice constant.

\section{Conclusion}\label{sec6}

We have studied using the Korringa-Kohn-Rostoker method the
appearance of single impurities and atomic swaps in the
half-metallic CrAs, CrSe and CrSb alloy crystallizing in the
zinc-blende structure. Although it was found that Cr antisites at
sp sites lead to half-metallic ferrimagnetism, we found that the
situation is much more complex. Cr single impurities at sp or
Void2 (see figure \ref{figa} for the structure) sites can couple
either antiferromagnetically (which is the energetically favored)
or ferromagnetically (in case of large lattice constants) to their
nearest Cr neighbors. Although in most cases the gap survives
there are cases like the case of Cr impurities at Void2 sites in
CrSe where the $e_g$ minority-spin states are pinned exactly at
the Fermi level.  Cr impurities at Void1 sites lead to a shift of
the unoccupied minority $e_g$ states lower in energy and when the
Fermi level is near the right edge of the gap in the perfect
compound the half-metallicity is lost. sp impurities at Cr, Void1
or Void2 sites lead to a redistribution of the p charge and they
are likely to induce new impurity states at the Fermi level. The
appearance of void impurities at Cr and sp atoms is the only case
where the gap is not affected at all. Finally we studied also the
case of atomic swaps between nearest neighbors. We found that in
the case of Cr-sp and Cr-Void2 swaps, the presence of the sp
impurity atom leads to new minority states pinned at the Fermi
level even in the cases where the half-metallicity was preserved
for single Cr impurities. Void1-sp swaps show behavior similar to
the single sp impurities at Void1 sites.

 Although impurity states are localized in space around the impurity atoms and very
fast we regain the bulk behavior, their interaction can lead to
wide impurity bands within the gap and thus the loss of the
half-metallic character. Only void impurities at Cr and sp sites
and Cr impurities at sp sites (as long as no swap occurs) keep the
half-metallic character of the perfect alloys. Our results suggest
that for CrAs alloys in the zinc-blende structure to be
implemented in spintronic devices, the prevention of the defects
creation is imperative.

%--------------------------------------------------------------------------------------------------------------------------

\end{document}